\documentclass[prb,reprint,superscriptaddress]{revtex4-1}
\usepackage{graphicx}
\usepackage{amssymb}
\usepackage{amsmath}
\usepackage{url}
\usepackage{prettyref}
	\newcommand{\pr}[1]{\prettyref{#1}}
	\newrefformat{app}{appendix~\ref{#1}}
	\newrefformat{eqn}{Eqn.~(\ref{#1})}
	\newrefformat{eqns}{Eqns.~(\ref{#1})}
	\newrefformat{sec}{section~\ref{#1}}
	\newrefformat{subsec}{section~\ref{#1}}
	\newrefformat{subsubsec}{section~\ref{#1}}
	\newrefformat{fig}{Fig.~\ref{#1}}
\usepackage[breaklinks,bookmarksnumbered]{hyperref}
\newcommand{\eps}{\epsilon}
\newcommand{\V}[1]{\mathbf {#1}}
\newcommand{\FullInterlayer}[2]{
\raisebox{#1}{
	\includegraphics[width=#2]{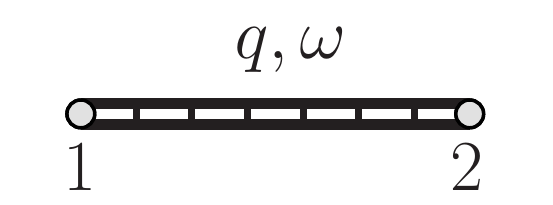}
}
}
\newcommand{\BarePhononD}[2]{
\raisebox{#1}{
	\includegraphics[width=#2]{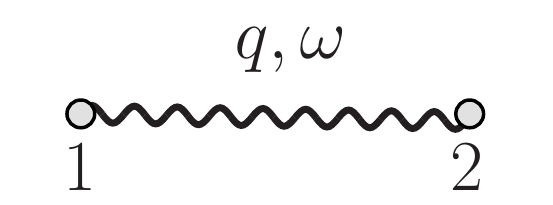}
}
}
\newcommand{\BareCoulombD}[2]{
\raisebox{#1}{
	\includegraphics[width=#2]{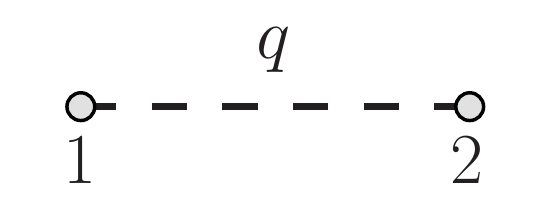}
}
}
\newcommand{\DysonInterlayerA}[2]{
\raisebox{#1}{
	\includegraphics[width=#2]{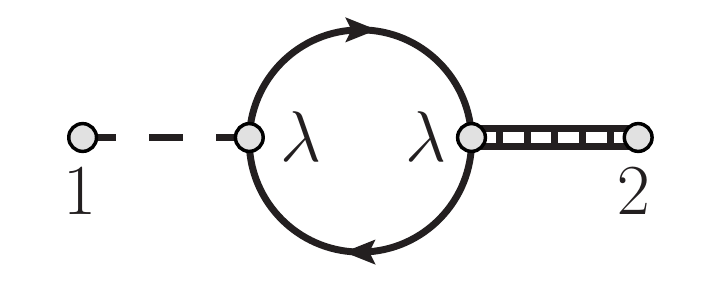}
}
}
\newcommand{\DysonInterlayerB}[2]{
\raisebox{#1}{
	\includegraphics[width=#2]{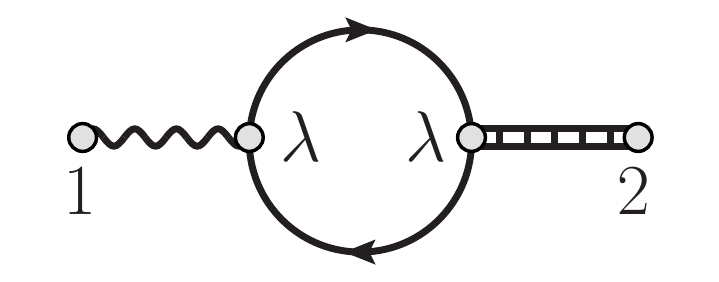}
}
}
\begin{document}
\title{Coulomb drag in graphene -- boron nitride heterostructures:
the effect of virtual phonon exchange}
\author{Bruno Amorim}
\affiliation{%
Instituto de Ciencia de Materiales de Madrid,
CSIC,
Cantoblanco,
E-28\,049, Madrid, Spain
}
\author{J{\"u}rgen Schiefele}
\affiliation{%
Departamento de F\'isica de Materiales,
Universidad Complutense de Madrid,
E-28\,040, Madrid, Spain%
}
\author{Fernando Sols}
\affiliation{%
Departamento de F\'isica de Materiales,
Universidad Complutense de Madrid,
E-28\,040, Madrid, Spain%
}
\author{Francisco Guinea}
\affiliation{%
Instituto de Ciencia de Materiales de Madrid,
CSIC,
Cantoblanco,
E-28\,049, Madrid, Spain
}
%
%
%
\date{September 28, 2012}
%
%
%
%
%
%
\begin{abstract}
For a system of two spatially separated monoatomic graphene
layers encapsulated in hexagonal boron nitride, 
we consider the drag effect between charge carriers in the Fermi liquid regime.
Commonly, the phenomenon is described in terms of an interlayer Coulomb interaction.
We show that if an additional electron -- electron interaction via exchange of virtual substrate phonons
is included in the model, the predicted drag resistivity 
is modified considerably at temperatures above 150\,K.
The anisotropic crystal structure  of boron nitride, with strong intralayer and 
comparatively weak interlayer bonds, is found to play an important role in this effect.
\end{abstract}
\maketitle
%
%
%
%
%
%
%
%
%
\section{Introduction}
If two systems containing mobile charge carriers are spatially separated
such that direct charge transfer is not possible, but close enough
to allow interaction between the carriers in different layers,
the resulting momentum transfer will equalize the drift velocities in
both systems.
This frictional effect
was experimentally observed between (quasi) two-dimensional electron gases in double quantum well
structures \cite{Gramila_1991,Sivan_1992}.
In most of the theoretical work the interlayer interaction was attributed to
Coulomb scattering, hence the effect now bears the name `Coulomb drag' 
(see Refs.~\onlinecite{Zheng_1993,Flensberg_1995,Kamenev_1995}).
Interest in the subject has been revived recently by the experimental progress
which made it possible to prepare two-dimensional electron systems based on monolayer graphene.
A considerable number of theoretical works 
\cite{Amorim_2012,Carrega_2012,Katsnelson_2011,Narozhny_2011,Peres_2011b,Hwang_2011,Wang_2007b,Badalyan2012,Benedikt2012,Ostrovski2012b,Levitov2012}
studied Coulomb drag between massless Dirac fermions, which effectively describe
the charge carriers in graphene \cite{Guinea_2009}.
However, a quantitatively correct explanation of the  experimental data 
is still lacking\cite{Amorim_2012,Kim_2011,Kim_2012,Gorbachev_2012}.
In the typical experiment, Coulomb drag is studied by driving a constant current $I_2$ 
through one of the layers (the active one, labeled by the index $\lambda=2$ in \pr{fig:layer_general}).
If no current is allowed to flow in the other (passive, index 1) layer, 
a potential difference $V_{1}$ builds up there.
In terms of these two quantities, the drag resistivity
$\rho_D \equiv (W/L) V_{1} / I_{2}$ serves as a measure of the momentum 
transfer between the two layers, where $W$ and $L$ are, respectively, the width and the length of the layer.
A theoretical expression for $\rho_D$ in second order in the interlayer interaction
can be derived either using Boltzmann's kinetic  equation\cite{Hwang_2011,Peres_2011b,Flensberg_1995,Jauho_1993}
or the Kubo formula \cite{Narozhny_2011,Hwang_2011,Kamenev_1995}.

In the present work, we focus on the interlayer interaction 
responsible for the drag effect in heterostructures composed of 
two graphene monolayers and hexagonal boron nitride (hBN),
see \pr{fig:layer_general}.
The large bandgap insulator hBN has a layered structure composed of stacked hexagonal crystal planes. 
Recently the material received much attention as it allows the construction of 
graphene -- hBN devices with, 
in comparison to the much used SiO$_2$ substrates,
favorable high carrier mobilities\cite{Dean_2010,Mayorov_2011,Schiefele_2012,Garcia_2012}.
In particular, the Manchester group reported the fabrication of devices 
where a few layer thin hBN crystal, obtained by exfoliation, is sandwiched between
two monolayers of graphene
\cite{Ponomarenko_2011,Britnell_2012,Britnell_2012b}.
If such a structure is used for a Coulomb drag experiment, the Dirac fermions in the
active and passive layer can exchange momentum not only via Coulomb interaction
but also by phonon exchange through the spacer medium.
The effect of a combined Coulomb-phonon coupling on the drag resistivity
has previously only been studied for quasi two dimensional electron gases in
semiconductor systems 
\cite{Jalabert_1989,Tso_1992,Zhang_1993c,Gramila_1993,Guven_1997,Bonsager_1998}.

In the following, we first investigate the effects of the anisotropy of hBN,
where the bonds in between the graphene-like planes are much weaker than the in-plane bonds,
on the electron -- electron interaction via phonon exchange.
We then show that the inclusion of phonon exchange into the description 
of Coulomb drag can significantly alter the temperature, density and distance  dependence of the predicted value for
$\rho_D$ at temperatures above 150\,K.
%
%
%
%
%
%
%
\begin{figure}
\centering
\includegraphics[width=0.8\columnwidth]{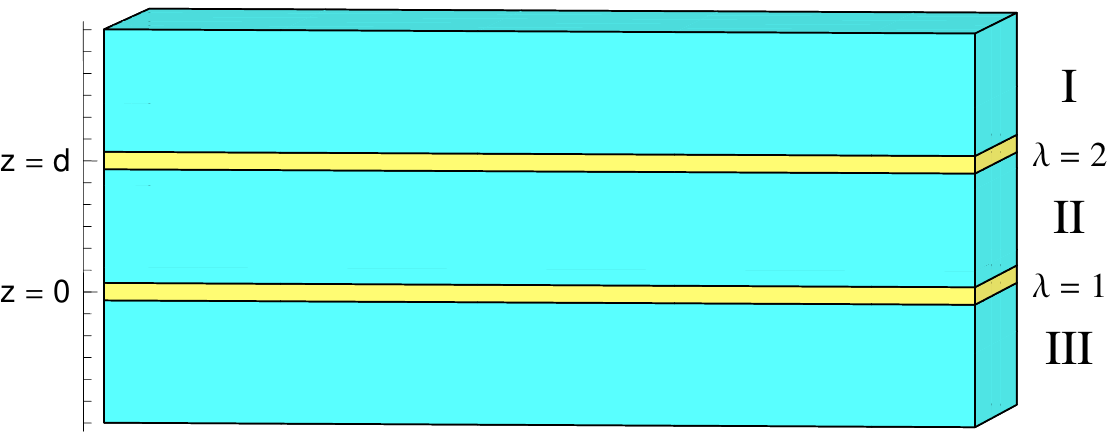}
\caption{%
A sketch of the double layer system under consideration.
The two monoatomic graphene layers (yellow) with charge carrier concentration 
$n_1$, $n_2$ are placed at $z=0$ and $z=d$ and labelled by the layer index
$\lambda = 1,2$, respectively.
The surrounding space (regions I, {II}, and {III}) is filled with the insulating material
boron nitride with hexagonal structure (hBN).
}
\label{fig:layer_general}
\end{figure}
%
%
%
%
\section{Interlayer interaction}
%
%
%
%
\subsection{Combined Coulomb -- phonon mediated interaction}
In a two-layer system as shown in \pr{fig:layer_general}, where the regions I, {II} and {III}
are filled with a homogeneous isotropic dielectric medium, 
the Fourier transform of the bare (unscreened) Coulomb potential  between electrons in layers $\lambda$ and $\lambda'$
has the form
\begin{align}
V_{\lambda \lambda'}^{(0)}(q) 
	&=
	\frac{1}{\eps_\infty}
	\frac{e^2}{2 \eps_{\rm vac}  q} e^{-q d (1 -\delta_{\lambda \lambda'})}
\;,
\label{eqn:V0}
\end{align}
where $\V q = (q_x, q_y)$,
$\eps_{\rm vac}$ denotes the dielectric constant of vacuum and  
$\eps_\infty$ accounts for the high frequency screening properties of the medium.
Apart from this Coulomb interaction, the charge carriers in each graphene layer 
interact via a substrate phonon mediated interaction. The charge carriers from each layer couple to the long range electric fields generated by optically active phonon modes in the surrounding material via Fr\"ohlich  coupling\cite{Frohlich_1954,Mahan_old,Marder}.
This remote interaction between carriers in graphene
and optical phonon modes  in a substrate medium was found to influence 
the electrical conductivity of graphene on a dielectric substrate\cite{Fratini_2008,Chen_2008,Schiefele_2012}.

In Appendix~\ref{app:Frolich}, we show that in an isotropic medium,  the combined interaction 
between electrons in layers $\lambda$ and $\lambda'$
via the effects of a static Coulomb potential and virtual substrate phonon exchange 
is of the form  of \pr{eqn:V0}, with $\eps_\infty$ replaced by the frequency dependent dielectric function $\eps(\omega)$
of the substrate material (see \pr{eqn:U0}).

In the following, we specialize to the anisotropic spacer material hBN.
From its three acoustic and nine optical phonon bands,
only those that (via dipole oscillations) create long range electric fields couple to the graphene electrons%
\footnote{%
See Refs.~\protect\onlinecite{Geick_1966, Michel_2011a, Serrano_2007} for details on the phonon dispersions of hBN,
and the classification of the vibrational modes into Raman active, infrared active and optically silent.
Figure~3 and eqns.~(21) and (24) of Ref.~\protect\onlinecite{Michel_2011a} show how the long range Coulomb
potential associated with the infrared active modes leads to the splitting of transverse and longitudinal optical
frequencies at the $\Gamma$ point.
}.
Given the layered uniaxial crystal structure of hBN, these (infrared active) optical  modes are
described by a dielectric tensor of the form\cite{Loudon}
\begin{equation}
\boldsymbol{\epsilon}(\omega)=\textrm{diag}\left[\epsilon_{\bot}(\omega),\epsilon_{\bot}(\omega),\epsilon_{\Vert}(\omega)\right]
\;.
\label{eqn:e_tensor}
\end{equation}
The resonance frequencies $\omega_{\rm TO}^{\Vert}$ and  $\omega_{\rm TO}^{\bot}$ of the two retarded\footnote{%
We are here using the retarded expression (defined as being analytic in the upper half 
of the complex $\omega$ plane) in order to be consistent with the likewise retarded 
polarizability of graphene taken from Ref.~\protect\onlinecite{Wunsch_2006}. 
Not keeping this consistency yields significantly different results.
} 
dielectric functions
\begin{equation}
\epsilon_{\bot,\Vert} (\omega)
	=
	\epsilon_{\infty}^{\bot,\Vert}
	+
	f_{\bot,\Vert}
	\frac
	{\bigl(\omega_{\rm TO}^{\bot,\Vert}\bigr)^{2}}
	{\bigl(\omega_{\rm TO}^{\bot,\Vert}\bigr)^{2}-\omega^{2}-i\omega \gamma_{\bot,\Vert}}
\;,
\label{eqn:dielectric}
\end{equation}
are the phonon frequencies at the $\Gamma$ point for transverse intraplane shear modes with displacements
parallel and perpendicular to the $c$-axis of the crystal 
(aligned with the $z$ direction in \pr{fig:layer_general}), respectively. We make the usual approximation of dispersionless optical phonon bands.\cite{Frohlich_1954, Das_Sarma_1985, Jalabert_1989}
The values for the high frequency dielectric constants $\epsilon_{\infty}$, 
the oscillator strengths $f$ (related to the static, $\epsilon_{0}$, and high frequency dielectric constants, $f=\epsilon_{0}-\epsilon_{\infty}$), 
$\omega_{\rm TO}$ and the damping factors $\gamma$ taken from Ref.~\onlinecite{Geick_1966} are listed in
Table~\ref{table_dielectric}.

%
%

To obtain
the  combined Coulomb-phonon interaction $U^{(0)}_{\lambda \lambda'}$ in the anisotropic medium,
we solve Poisson's equation 
\begin{equation}
-\nabla \cdot (\boldsymbol{\epsilon}\cdot\nabla\phi)=\rho_{\rm free}/\epsilon_{\rm vac} 
\nonumber
\end{equation}
with $\rho_{\rm free}$ being the free charge density of a point charge $-e$ at the origin. 
With \pr{eqn:e_tensor}, Poisson's equation becomes
\begin{equation}
-\frac{\partial}{\partial z}\left(\epsilon_{\Vert}
\frac{\partial}{\partial z}\phi(\boldsymbol{q},z)\right)+q^{2}\epsilon_{\bot}\phi(\boldsymbol{q},z)
=-\frac{e}{\epsilon_{{\rm vac}}}\delta(z)
\,,
\nonumber
\end{equation}
%
and as $U^{(0)}_{12}=-e\phi(q,d)$ and  $U^{(0)}_{11}=U^{(0)}_{22}=-e\phi(q,0)$ we get 
\begin{align}
U_{\lambda \lambda'}^{(0)}(q, \omega) 
	&=
	\frac{e^2}{2 \eps_{\rm vac}  {\eps_{\Vert}(\omega)}q}\sqrt{\frac{ {\eps_{\Vert}(\omega)}}{{\eps_{\bot}(\omega)}}} 
\label{eqn:U0_hBN}
\\
	&\times
	\operatorname{exp}
	\biggl[-q d (1 -\delta_{\lambda \lambda'}) \sqrt{\frac{\eps_{\bot}(\omega)}{\eps_{\Vert}(\omega)}} \biggr]
\;.
\nonumber
\end{align}
A generalization of this result to structures where the regions I,{II}, and {III}
(see \pr{fig:layer_general}) are filled with different insulating materials
(or air)  is straightforward; 
$U_{1 1}^{(0)}$ then involves different dielectric functions than $U_{2 2}^{(0)}$.

%
%
%
%
\subsection{RPA screened interlayer interaction}
To take into account the screening properties of the conduction electrons in the graphene layers themselves,
we employ the standard procedure of solving the Dyson equation for the two-layer system within
the random phase approximation (RPA) (see Ref.~\onlinecite{Kamenev_1995}).
This finally yields the dressed interlayer interaction
\begin{align}
U_{1 2} (q,\omega)
	&=
	\frac 
	{U^{(0)}_{1 2}(q,\omega)}
	{\eps_{\rm RPA}(q,\omega)}
\;.
\label{eqn:interlayer}
\end{align}
The total screening function for the coupled electron-phonon system given by 
(see Ref.~\onlinecite{Zhang_1993c} and Appendix \ref{app:mathematical})
\begin{align}
\eps_{\rm RPA}
	&=
	(1- U^{(0)}_{11} \chi_1 ) 
	(1- U^{(0)}_{22} \chi_2 ) 
	- U^{(0)}_{12} U^{(0)}_{21} \chi_1 \chi_2
\;,
\label{eqn:epsRPA}
\end{align}
where $\chi_{1,2}$ denotes the (frequency and momentum dependent) polarizability
of the graphene layers\footnote{%
In the numerical calculations, we use for simplicity the zero temperature 
expression for $\chi$ as calculated in Refs.~\protect\onlinecite{Wunsch_2006,Hwang_2007},
which is a good approximation for $T \ll T_F$,
with $T_F$ the Fermi temperature.}.
%

%
%
%
%
%
%
%
%
%
\begin{figure}
\centering
\includegraphics[width=0.8\columnwidth]{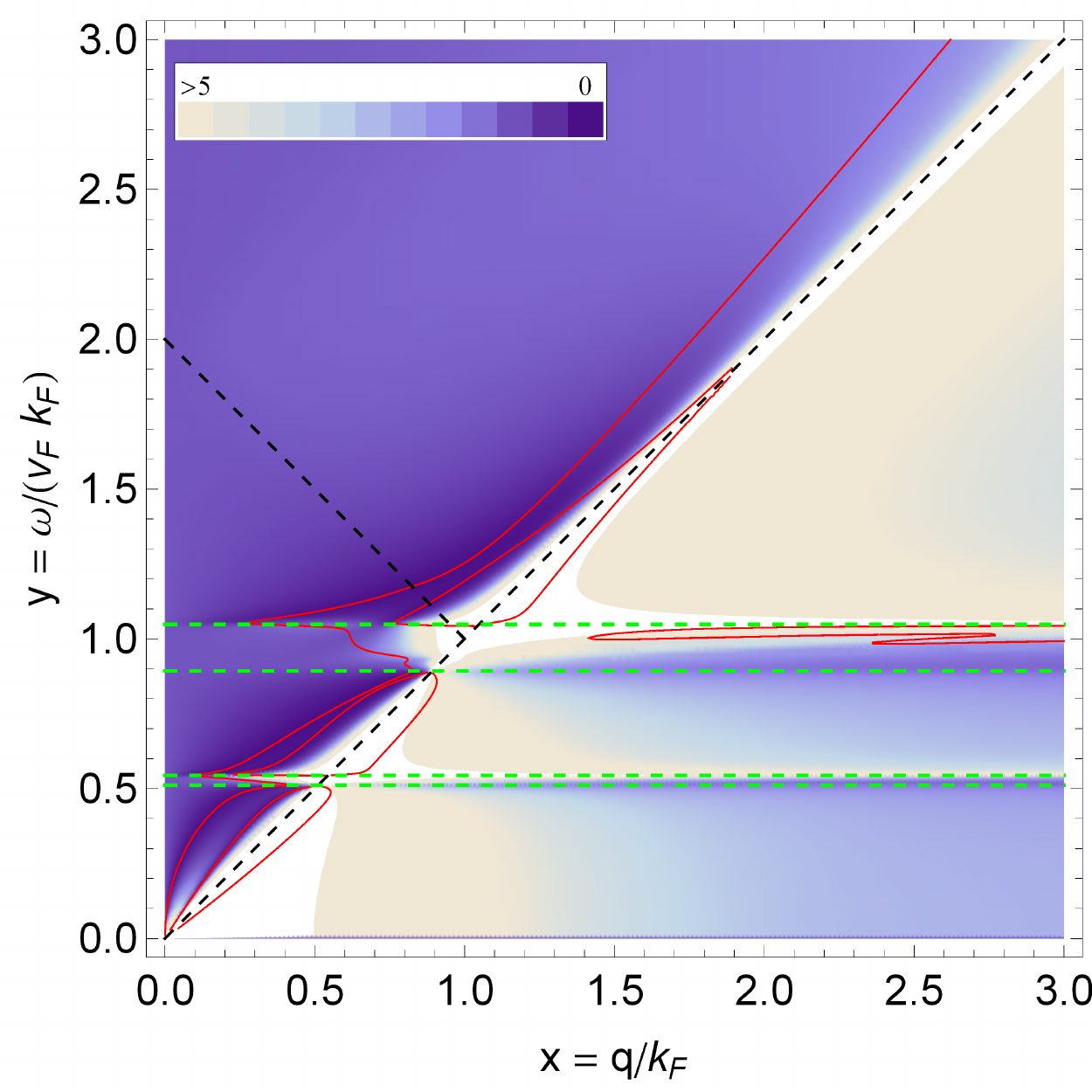}
\caption{%
Absolute value of the total screening function $\eps_{\rm RPA}$ \pr{eqn:epsRPA},
with $n_1=n_2=0.02$\,nm$^{-2}$ and $d=8$\,nm.
Vertical green lines show the optical resonance frequencies 
$\omega_{\rm TO}^{\Vert}$,
$\omega_{\rm LO}^{\Vert}$,
$\omega_{\rm TO}^{\bot}$, and
$\omega_{\rm LO}^{\bot}$
of hBN (bottom to top).
Red curves mark the zeros of $\operatorname{Re} {\eps_{\rm RPA}}$. 
The dashed black lines show the line $y=x$ and mark the region where ${\rm Im}\chi=0$. The hybridization between  phonon and plasmon modes is clear.
}
\label{fig:epsRPA}
\end{figure}
%
\pr{fig:epsRPA} shows a density plot of $|\eps_{\rm RPA}(q,\omega)|$,
using dimensionless units $x=q/k_{F}$ and $y=\omega/(v_{F}k_{F})$, where $k_{F}$ is the Fermi momentum.
The horizontal dashed green lines mark the transverse and longitudinal frequencies
of the infrared active modes in hBN, connected by the Lyddane-Sachs-Teller
relation\cite{Lyddane_1941} $\omega_{\rm LO}^{2}/\omega_{\rm TO}^{2}=\epsilon_{0}/\epsilon_{\infty}$.
For small damping $\gamma \ll \omega_{\rm TO}$, the real parts of 
$\epsilon_{\bot,\Vert} (\omega)$ are close to a pole at $\omega_{\rm TO}^{\bot,\Vert}$ and
close to zero at $\omega_{\rm LO}^{\bot,\Vert}$, respectively.
Near these frequencies, the absolute value of the total screening function $\eps_{\rm RPA}$ likewise
shows an abrupt change from high values (light colors) to almost zero (dark colors).
In regions where $\left\vert \eps_{\rm RPA} \right\vert$ is small, the red lines 
$\operatorname{Re} \eps_{\rm RPA} = 0$
show the coupled plasmon-phonon dispersion relation of the two-layer system.
%
%
%
%
%
%
%
%
%
%
%
%
%
%
\section{Results for the drag resistivity}
%
%
%
%
%
%
%
%
%
%
%
\begin{figure}
\centering
\includegraphics[width=0.8\columnwidth]{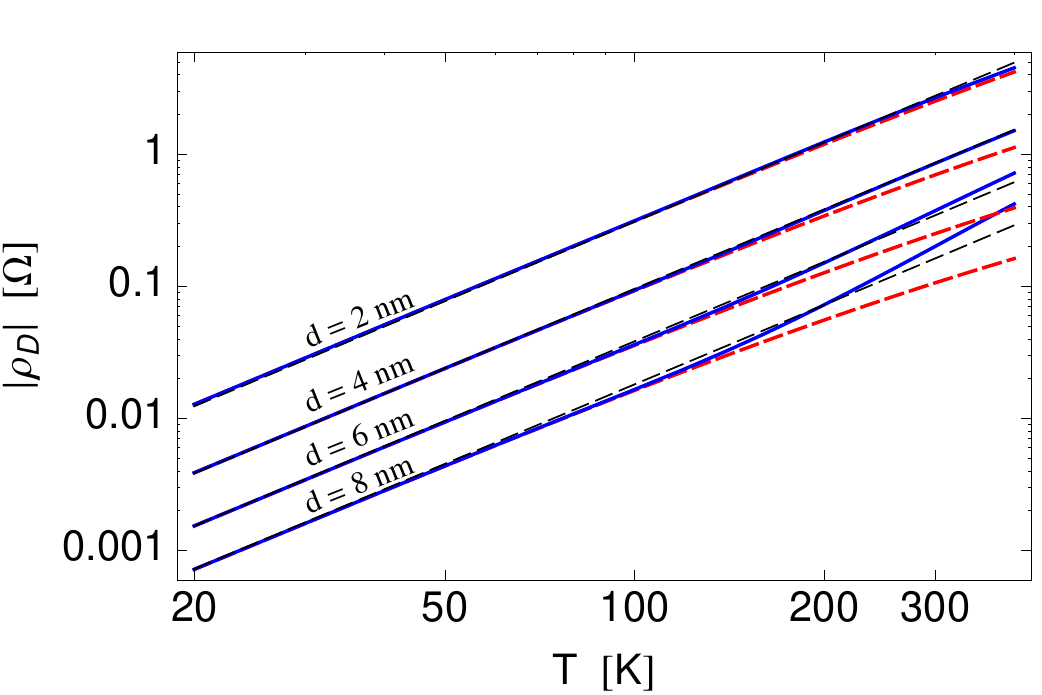}
\caption{%
Drag resistivity   versus temperature for various interlayer
distances, $n=0.02$\,nm$^{-2}$.
The blue curves show $|\rho_{\rm D}|$ (\pr{eqn:rho_general}) including interaction via phonon 
exchange and Coulomb interaction,
the dashed red curves show $|\rho_{\rm CD}|$ (\pr{eqn:Coulomb_drag}) with Coulomb 
interaction  only, and 
dashed black lines the low-temperature asymptote $\rho_{\rm CD}^{\rm low \, T}$ (\pr{eqn:low_T}). 
The lowest pair of curves ($d=8$\,nm) is also plotted on a linear scale in \pr{fig:Ani_vs_Iso}.
}
\label{fig:rho_vs_T}
\end{figure}
%
%
%
%
%
%
%
%
%
\begin{figure}
\centering
\includegraphics[width=0.8\columnwidth]{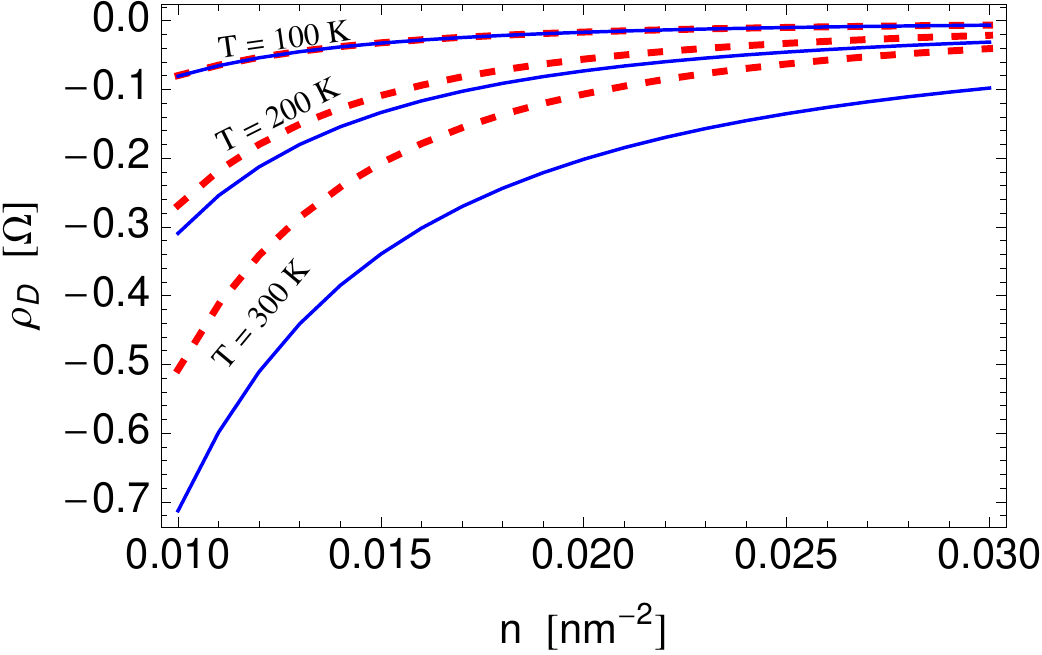}
\caption{%
Drag resistivity versus carrier density for various temperatures, 
$d=8$\,nm.
Colors as in \pr{fig:rho_vs_T}.
}
\label{fig:rho_vs_n}
\end{figure}
%
%
%
%
%
%
%
%
%
\begin{figure}
\centering
\includegraphics[width=0.8\columnwidth]{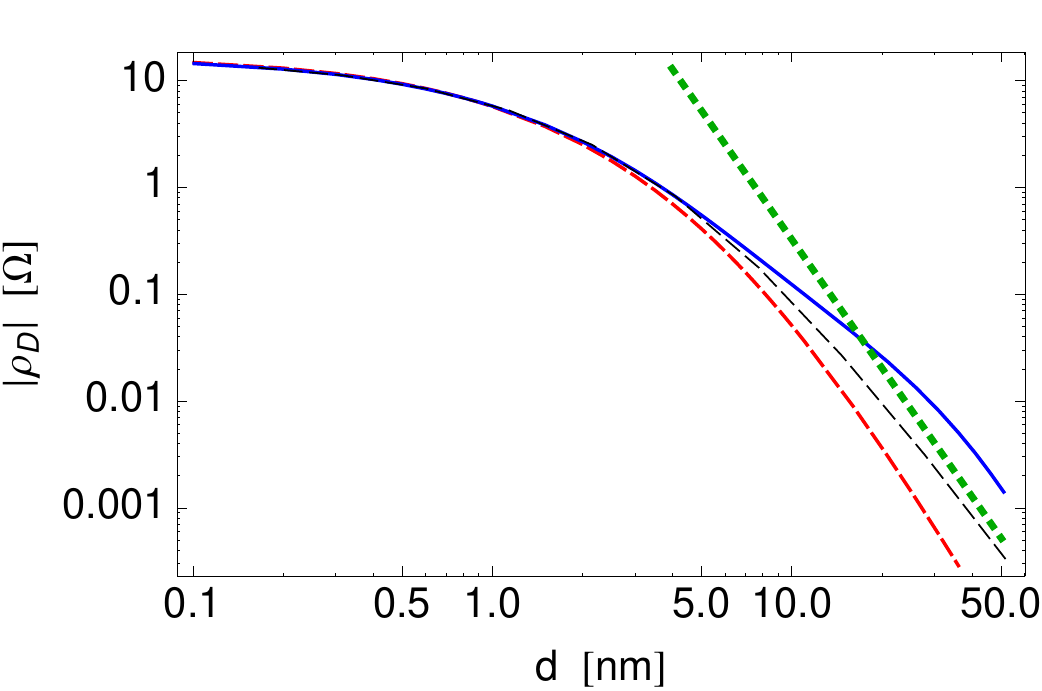}
\caption{%
Drag resistivity versus layer separation for $T=300$\,K, 
$n=0.02$\,nm$^{-2}$.
Colors as in \pr{fig:rho_vs_T}.
The low-temperature asymptote $\rho_{\rm CD}^{\rm low \, T}$ (dashed black curve, \pr{eqn:low_T}),
converges for large layer separation to $\rho_{\rm CD}^{\rm large \; d}$
(dotted green line, \pr{eqn:low_T_large_d}). Note that at this temperature and density $\rho_{\rm CD}^{\rm low \, T}$ already differs from the full static calculation, $\rho_{\rm CD}$.
}
\label{fig:rho_vs_d}
\end{figure}
%
%
%
%
%
%
%
\begin{figure}
\centering
\includegraphics[width=0.8\columnwidth]{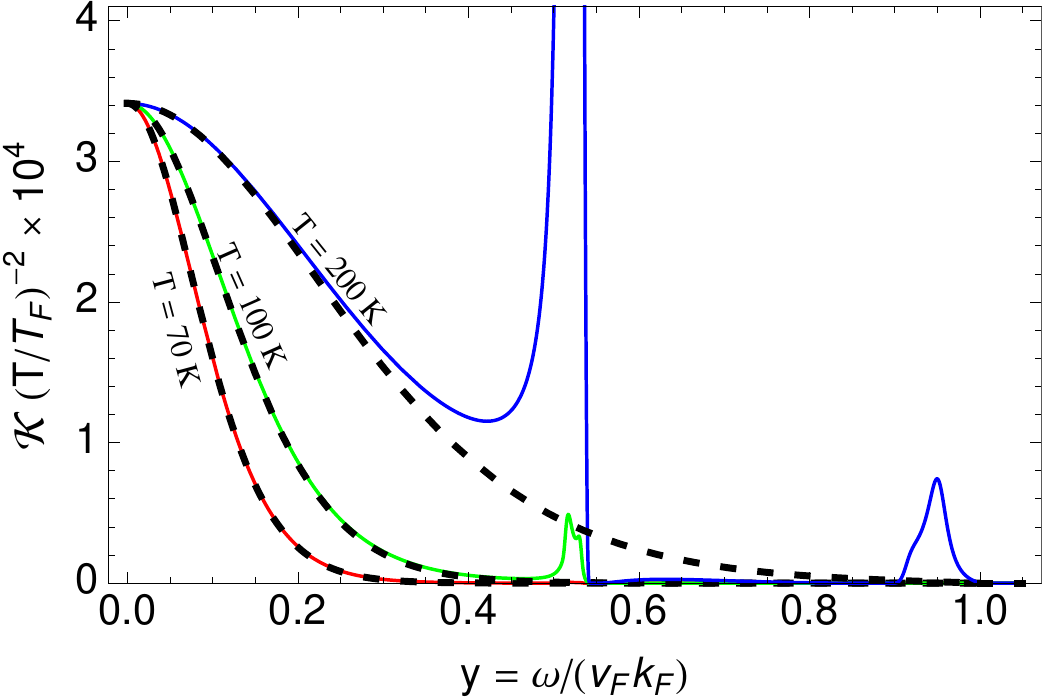}
\caption{%
The integral kernel 
$\mathcal{K}$ of \pr{eqn:kernel}  with $x=q/k_{F}=1$,
$d=8$\,nm, $n=0.02$\,nm$^{-2}$ as a function of $y=\omega/v_F k_F$ for the temperatures
200, 100, 70\,K (full curves, from top to bottom).
The curves have been aligned on the left side by dividing  with  the $y\to0$ temperature 
dependence  $(T/T_{\rm F})^{2}$.
At $T=100$\,K (green curve) a peak near the resonance frequency 
$\omega_{\rm TO}^{\Vert}$ appears, at $T=200$\,K (blue curve) there is and additional second peak near $\omega_{\rm TO}^{\bot}$ 
(see the vertical green lines in \pr{fig:epsRPA}).
Dashed curves show the integral kernel for $\rho_{\rm CD}$, where these peaks are absent.
}
\label{fig:kernel}
\end{figure}
%
%
%
%
%
%
%
%
%
In the following, we assume for the sake of simplicity 
the same positive carrier density $n$ (corresponding to electron doping)
in both layers, such that $E_F \gg k_{\rm B} T$.
In particular, we do not address the recently reported 
drag at charge neutrality point\cite{Gorbachev_2012},
which was attributed either to contributions from higher order perturbation theory\cite{Ostrovski2012b} 
or to correlated density inhomogenities in the graphene layers\cite{Levitov2012,Gorbachev_2012}.

The drag resistivity then assumes a negative value\cite{Kim_2011},
and the first non-vanishing contribution to $\rho_D$ obtained in perturbation theory 
is of second order in the dressed interlayer interaction\cite{Flensberg_1995,Kamenev_1995}.
In terms of the variables carrier density, layer separation, and
temperature, and under the assumptions that both layers are with high electron doping 
and $T \ll T_F$%
\footnote{%
We here use a simplified form of the nonlinear susceptibility of graphene, which is valid for electron
doping high enough such that the existence of the valence band can be ignored.
The condition $T\ll T_F$  is important as we use the zero temperature expressions for the polarizability 
of graphene.
See Refs.~\protect\onlinecite{Peres_2011b,Amorim_2012} for a discussion of both approximations.
},
it reads (refer to Refs.~\onlinecite{Peres_2011b,Amorim_2012} for details)
\begin{align}
\rho_D
	&=
	-\frac{\hbar}{e^2}
	\frac{\alpha_g^2}{8}
	\frac{\hbar v_F \sqrt{\pi n}} {k_B T}
	\int_0^\infty dx 
	\int_0^\infty dy 
	\;\mathcal{K} (T,d,n)
\;,
\label{eqn:rho_general}
\end{align}
where
$\alpha_g = e^2/(4 \pi \eps_{\rm vac} v_F \hbar)$
denotes the effective fine structure constant in graphene and
the integral kernel
\begin{align}
\mathcal{K}
	&=
	\frac{k_{F}^{2}\epsilon^{2}_{\rm vac}}{e^4}
	\frac{|U_{1 2}(x,y)|^2}{\operatorname{sinh}^2 \bigl( y \frac{T_F}{2 T} \bigr)}
	\frac{x^7 \,\Phi^2(x,y)}{x^{2}- y^{2}}
\;.
\label{eqn:kernel}
\end{align}
The function $\Phi$, defined in  \pr{eqn:phi_def}, is related to the nonlinear susceptibility of graphene,
and restricts the integration range in the $x,y$-plane to the region $\omega < v_F q$.

In order to estimate the contribution of phonon exchange to the drag effect,
we note that the drag resistivity $\rho_{\rm CD}$
resulting from Coulomb interaction only (which is usually taken as a measure for
Coulomb drag) is obtained by substituting the static value of the electron-electron interaction
into the integral kernel \pr{eqn:kernel}:
\begin{align}
\rho_{\rm CD}
	&=
	\rho_D \big|_{U^{(0)}_{\lambda \lambda^{\prime}} (q,\omega=0)}
\;.
\label{eqn:Coulomb_drag}
\end{align}
For low temperatures $E_{F} \gg k_{\rm B} T$, the resistivity $\rho_{\rm CD}$ can be approximated by 
$\rho_{\rm CD}^{\rm low \, T} \propto T^2$ of \pr{eqn:low_T} (see Ref.~\onlinecite{Amorim_2012}
for a detailed derivation),
under the additional condition $k_F d, k_F d/\epsilon_{\Vert} \gg 1$ (large layer spacing), this 
can be further approximated to yield\cite{Amorim_2012}
\begin{align}
\rho_{\rm CD}^{\rm large \; d}
	&=
	-\frac{\hbar}{e^2}
	\frac{\bigl(\eps_0^\Vert \bigr)^3}{\eps_0^\bot}
	\frac{\zeta(3)}{\pi 2^8 \alpha_g^2}
	\frac{(k_B T)^2}{(\hbar v_F)^2 n^3 d^4}
\;.
\label{eqn:low_T_large_d}
\end{align}
(Note that in the static limit,  one only needs to rescale 
$d\rightarrow d\sqrt{\epsilon_{\bot}/\epsilon_{\Vert}}$ and 
$\alpha_{g}\rightarrow\alpha_{g}/\sqrt{\epsilon_{\bot}\epsilon_{\Vert}}$
to take into account the anisotropy of hBN.)
The full blue curves in Figures \ref{fig:rho_vs_T}
-\ref{fig:rho_vs_d} show the absolute value of $\rho_D$ \pr{eqn:rho_general} for different parameters
$T$, $n$, and $d$,
while $|\rho_{\rm CD}|$ is shown by dashed red curves,
$\rho_{\rm CD}^{\rm low \, T}$ by the dashed black lines in Figures \ref{fig:rho_vs_T} and \ref{fig:rho_vs_d},
and the dotted green line in \pr{fig:rho_vs_d} shows $\rho_{\rm CD}^{\rm large \; d}$.

As Figures \ref{fig:rho_vs_T} and \ref{fig:rho_vs_n} show, the contribution of phonon  mediated interaction
to the drag resistivity is vanishingly small at low temperatures,
but becomes noticeable for $T>150$\,K, the effect being more pronounced the larger the layer separation.
This temperature dependence is due the factor
$\operatorname{sinh}^{-2} [ y T_F/(2 T) ]$
in the integration kernel \pr{eqn:kernel}, which suppresses the integrand for values
of $y > T/T_F$.
Thus at low temperatures, the main contribution to the $y$-integration in \pr{eqn:rho_general} 
comes from a frequency range where the dielectric functions in the integrand are still close
to their static values. 
However, the phonon contribution becomes noticeable at lower temperatures than one would expect,
taking into account that the energy of the lowest phonon mode $\hbar\omega_{\rm TO}^\Vert / k_B  \approx 1100\,K$. 
It is also  interesting to notice that in the range from 100 to 250\,K, the drag resistivity $\rho_{D}$,
including the effect of phonons,
is closer to the $T^2$ behaviour $\rho_{\rm CD}^{\rm low \, T}$ than the purely Coulomb drag result, 
$\rho_{\rm CD}$.  The  plot of $\mathcal{K}$  as a function of $y$ in \pr{fig:kernel} shows the origin of the phonon contribution
to the integral $\rho_D$:
With rising temperature,  peaks near the resonance frequencies 
$\omega_{\rm TO}^{\Vert}$ and  $\omega_{\rm TO}^{\bot}$ appear in the integrand,
which enhance the magnitude of $\rho_D$. 

\pr{fig:rho_vs_n} shows that the  relative effect of phonon exchange on $\rho_D$
is larger for high densities.
For high values of $n$, the argument of the  dielectric functions $\eps(\omega) = \eps(y v_F \sqrt{\pi n})$ 
in \pr{eqn:kernel} reaches the resonance frequency already at lower values of $y$. 
While $\rho_{\rm CD}$ decreases rapidly with $n$ due to increased screening of the Coulomb interaction,
the modification of the screening function $\eps_{\rm RPA}$ by phonon interaction is seen to  counteract this decrease
at high temperatures. 

\begin{figure}
\centering
\includegraphics[width=0.8\columnwidth]{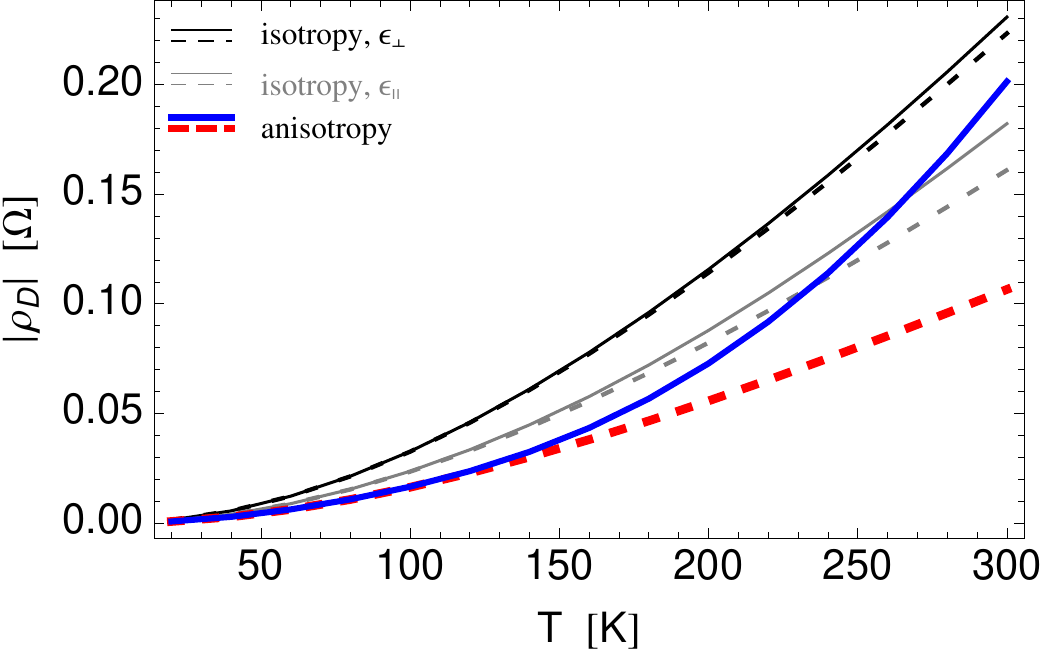}
\caption{%
Effect of the anisotropy of hBN on the behavior of drag with temperature,
with $n=0.02$\,nm$^{-2}$ and  $d=8$\,nm. 
The curves \textit{isotropy, $\epsilon_\Vert$} and \textit{isotropy, $\epsilon_\bot$} 
were computed assuming that the graphene layers are immersed in an isotropic dielectric medium, 
with dielectric functions given by $\epsilon_\Vert$ and $\epsilon_\bot$, respectively (see \pr{eqn:e_tensor}). 
The curve \textit{anisotropy} was computed taking into account the anisotropy of hBN as in \pr{eqn:U0_hBN}. 
Solid curves show $\rho_{\rm D}$, dashed ones $\rho_{\rm CD}$.
}
\label{fig:Ani_vs_Iso}
\end{figure}

Finally, \pr{fig:Ani_vs_Iso} illustrates the effect of 
the anisotropy in hBN that enters $\rho_D$ through the electron-electron interaction \pr{eqn:U0_hBN}. 
We compare the drag resistivity in hBN with that in an isotropic medium with dielectric functions
$\epsilon_\Vert$ and $\epsilon_\bot$, respectively (see \pr{eqn:e_tensor}).
The difference in magnitude between $\rho_D$ and $\rho_{\rm CD}$ is seen to be greatest in
the anisotropic case, where both in-plane and out-of-plane phonon modes contribute to
the interlayer interaction.
\begin{table}[h]
\caption{
Parameters for the dielectric function of hBN (see \pr{eqn:dielectric}) taken 
from Ref.~\onlinecite{Geick_1966}\footnote{%
The experimental data in Ref.~\protect\onlinecite{Geick_1966} 
exhibits two resonances, a strong and a weaker one, for each direction of the polarization of incident light. 
The weaker ones are attributed to  missorientation of the polycrystalline samples.
}.
}
\begin{ruledtabular}
\begin{tabular}{r | c c}
\, 		          & $\epsilon_{\bot}$  & $\epsilon_{\Vert}$	\\
\hline
$\epsilon_\infty$ &   4.95  		   & 4.10			\\	
$f$		          &   1.868			   & 0.532			\\
$\gamma$	      &   3.61\,meV		   & 0.995\,meV		\\
$\omega_{\rm TO}$	  &   170\,meV		   & 97.4\,meV		\\
\end{tabular}
\end{ruledtabular}
\label{table_dielectric}
\end{table}
%
%
%
%
%
%
\section{Summary and discussion}
We showed that including the electron -- electron interaction via phonon
exchange into the theory of Coulomb drag 
significantly changes the magnitude of the predicted 
drag resistivity in graphene-hBN heterostructures.
For large layer separations, the deviations become noticeable at temperatures higher than
150\,K. 

As the lowest phonon resonance frequency in the spacer material hBN corresponds to a temperature
of approximately 1100\,K,
our result at first sight seems to be at odds with the notion that phonon effects should
be proportional to the thermal population factor of the relevant modes.
This is indeed the case for other transport phenomena, 
like the substrate limited electron mobility in graphene,
where real momentum transfer from an electronic state
(in graphene) to a phonon mode (in a dielectric substrate material) plays a role\cite{Schiefele_2012,Fratini_2008}.
The decay rate of the electronic state is then overall proportional to the thermal population
of the phonon mode. Our scenario however involves the exchange of virtual phonons
in a process that is of second order in the interlayer interaction\cite{Kamenev_1995},
and no decay processes into real phonon states are relevant for $\rho_D$. 
We note that in Ref.~\onlinecite{Carrega_2012}, the effect of substrate phonons on Coulomb drag
was considered for the case where a material described by a uniform dielectric function fills
what is our region {III} of Fig.~\ref{fig:layer_general},
and a deviation from the low-temperature $T^2$ behavior of $\rho_D$ was
predicted for temperatures roughly an order of magnitude lower than the phononic resonance frequency of the substrate material.

Up to date, there remains considerable discrepancy between experimental data on Coulomb drag between 
graphene layers embedded in SiO$_2$/Al$_2$O$_3$\cite{Kim_2011,Kim_2012}
and  hBN\cite{Gorbachev_2012} and the existing theoretical work.
For hBN, the reported drag resistivities in the Fermi liquid regime are roughly a factor of three larger 
than predicted, and the results of the present paper do not change this situation.
The experimentally reported  $T^2$ dependence of $\rho_D$ for $d=6$\,nm and $n=0.018$\,nm$^{-2}$ 
up to temperatures of 240\,K\footnote{%
See Fig.~2a in Ref.~\protect\onlinecite{Gorbachev_2012}.}
does not disagree with our results presented in \pr{fig:rho_vs_T}. 
Actually, it appears that in the temperature range of 100 to 250\,K the inclusion 
of phonon mediated interaction brings the behaviour of drag closer to the low temperature $T^2$ behaviour 
than with static Coulomb interaction only. 
Nevertheless, an extension of the experimental data shown in Ref.~\onlinecite{Gorbachev_2012} up to room 
temperature would be needed to distinguish clearly between $\rho_D$ and $\rho_{\rm CD}$. 
The use of other substrate materials, such as SiO$_2$, should not qualitatively alter the results of this paper.    
We think that future experiments with devices as considered in the present
work will be able to check our predictions.
%
%
%
%
%
%
\begin{acknowledgements}
The authors would like to thank N.M.R. Peres for useful discussions.
Financial support from 
Funda\c{c}\~{a}o
para a Ci\^{e}ncia e a Tecnologia (Portugal) through Grant No. SFRH/BD/78987/2011 (B.A.),
the Marie Curie ITN \emph{NanoCTM} (J.S.)
and from MICINN (Spain) through Grant No. 
FIS2010-21372 (F.S.)
and 
FIS2008-00124 (F.G.)
is acknowleged.
\end{acknowledgements}
%
%
%
%
%
%
%
%
%
\appendix
%
%
%
\section{Fr\"ohlich electron--phonon coupling and phonon mediated electron--electron interaction}
\label{app:Frolich}
Throughout the present work, we 
assume the dielectric properties of hBN layers 
forming heterostructures as shown in Fig.~\ref{fig:layer_general} to be the same as for bulk hBN. 

The Fr\"ohlich Hamiltonian describing the coupling of electrons to a bulk polar longitudinal phonon mode 
(in an  isotropic homogeneous dielectric material) 
is given by\cite{Frohlich_1954,Mahan_old,Marder}
\begin{equation}
H_{e-ph}=\int d^{3}r\rho(\V r)\frac{1}{\sqrt{V}}\sum_{\V Q}M(Q)e^{i\V Q\cdot \V r}\left(a_{\V Q}-a_{-\V Q}^{\dagger}\right),
\nonumber
\end{equation}
where $\rho(\V r)$ denotes the electron density operator, $a_{\V Q}^{\dagger}$
($a_{\V Q}$) the creation (annihilation) phonon operator with momentum
$\V Q=(q_{x},q_{y},q_{z})$ and the matrix element reads 
\begin{equation}
M(Q)=i\sqrt{\frac{e^{2}\omega_{\rm LO}}{2\epsilon_{\rm vac}Q^{2}}\left(\frac{1}{\epsilon_{\infty}}-\frac{1}{\epsilon_{0}}\right)},
\end{equation}
%
with the longitudinal optical
phonon frequency $\omega_{\rm LO}$. 
The phonon mediated interaction between electrons
is given by
\begin{equation}
\psi(Q,\omega)=M(Q)M(Q)^{*}D_{\rm LO}(Q,\omega),
\label{eqn:phonon_app}
\end{equation} 
with the bare phonon propagator
\begin{equation}
D_{\rm LO}(Q,\omega)=2\omega_{\rm LO}/\left(\omega^{2}-\omega_{\rm LO}^{2}\right)
\;.
\end{equation} 
We employ the usual approximation of dispersionless optical phonons \cite{Frohlich_1954,Das_Sarma_1985,Jalabert_1989,Guven_1997}.

The bare Coulomb interaction is given by
\begin{equation}
V_{\rm C}=e^{2}/\left(\epsilon_{\rm vac}\epsilon_{\infty}Q^{2}\right),
\label{eqn:Coulomb_app}
\end{equation}
where $\epsilon_{\infty}$ takes into
account the high frequency screening properties of the medium.

With Eqns.~(\ref{eqn:phonon_app})-(\ref{eqn:Coulomb_app}) and the
Lyddane-Sachs-Teller relation\cite{Lyddane_1941} $\omega_{\rm LO}^{2}/\omega_{\rm TO}^{2}=\epsilon_{0}/\epsilon_{\infty}$,
we arrive at the combined Coulomb and phonon mediated interaction 
\begin{equation}
U(Q,\omega)  =  V_{\rm C}(Q)+\psi(Q,\omega)=\frac{e^{2}}{\epsilon_{\rm vac}\epsilon(\omega)Q^{2}},\label{eq:combined}
\end{equation}
%
with $\eps(\omega)$ the dielectric function of the medium, see Eqn.~\ref{eqn:dielectric}.   
Since the Fr\"ohlich coupling is derived in a phenomenological
approach based on the dielectric properties of the material,
the combined Coulomb and phonon mediated interaction
simply reduces to the Coulomb interaction screened by $\epsilon(\omega)$, as it should.

%
In a two-layer system as shown in \pr{fig:layer_general}, 
the Fr\"ohlich  coupling coupling between bulk phonons and 2D electrons of layer $\lambda$ is given by\cite{Das_Sarma_1985}
\begin{align}
	M_\lambda ( q, q_z)
	&=
	i \, \sqrt{
	\frac {e^2 \omega_{\rm LO}}{2 \eps_{\rm vac} (q^2 + q_z^2)}  
	\biggl( \frac{1}{\epsilon_\infty} - \frac{1}{\epsilon_0}\biggr)
	}
	e^{i q_z d (1-\delta_{\lambda 1})}
\;,
\label{eqn:Frohlich2D}
\end{align}
where $\V q = (q_x, q_y)$ is a two-dimensional momentum vector. 
In analogy to the  above,  we now get for the combined Coulomb--phonon interaction
in a homogeneous isotropic medium
\begin{align}
U_{\lambda \lambda'}^{{\rm iso}}(q, \omega) 
	&\equiv
	V_{\lambda \lambda'} (q) 
	+
	\psi_{\lambda \lambda'}    (q, \omega)
\nonumber
\\
	&=
	\frac{1}{\eps(\omega)}
	\frac{e^2}{2 \eps_{\rm vac}  q} e^{-q d (1 -\delta_{\lambda \lambda'})}
\;.
\label{eqn:U0}
\end{align}

Although it is possible to generalize the Fr\"ohlich electron--phonon coupling for the case of 
anisotropic materials\cite{Loudon,Stroscio} and inhomogeneous layered materials\cite{Mori}, 
the easiest way to obtain the effective electron--electron interaction, taking into account the phonon mediated interaction, is by solving Poisson's equation for the electric potential created by a point charge in the dielectric medium taking into account the frequency dependence of its dielectric tensor. 

%
%
\section{Mathematical details}
\label{app:mathematical}
The dressed interlayer interaction \pr{eqn:interlayer} is the solution of the coupled set of Dyson
equations
%
\label{eqn:Dyson}
\begin{align}
U_{1 2}&(q, \omega) 
	= \FullInterlayer{-1.8ex}{0.25\columnwidth}
\nonumber
\\
	&=
	\BareCoulombD{-1.8ex}{0.25\columnwidth}
	+
	\sum_{\lambda=1}^2
	\DysonInterlayerA{-3.1ex}{0.32\columnwidth}
\nonumber
\\
	&+
	\BarePhononD{-1.8ex}{0.25\columnwidth}
	+
	\sum_{\lambda=1}^2
	\DysonInterlayerB{-3.1ex}{0.32\columnwidth}
\;,
\nonumber
\end{align}
%
where the dashed and wiggled lines denote the bare Coulomb and
phonon interaction, respectively, and the full curves electron
propagators
(see Refs.~\onlinecite{Zhang_1993c,Kamenev_1995}).

The function $\Phi(x,y)$ appearing in \pr{eqn:kernel} reads \cite{Peres_2011b,Amorim_2012}
\begin{align}
\Phi(x,y)
	&=
	\Phi^+(x,y) \,\Theta(y-x+2) \Theta(x-y)
\nonumber
\\
	&+
	\Phi^-(x,y) \, \Theta(1-y-|1-x|)
\;,
\label{eqn:phi_def}
\end{align}
where
\begin{align}
\Phi^\pm
	&=
	\pm \operatorname{cosh}^{-1} \biggl( \frac{2\pm x} y \biggr)
	\mp
	\frac{2\pm x} y \sqrt{\biggl( \frac{2\pm x} y \biggr)^2 -1}
\;.
\nonumber
\end{align}

For the low-temperature approximation of $\rho_D$,
the factor $\operatorname{sinh}^{-2} [ y T_F/(2 T) ]$
in the integration kernel \pr{eqn:kernel}, which suppresses the integrand for values
of $y > T/T_F$,
allows one to expand the remaining integrand to the lowest order of $y$.
The $y$ integration can then be performed, yielding
\begin{align}
\rho_{\rm CD}^{\rm low \, T}
&	=
	-\frac{\hbar}{e^2}
	\frac {2 \pi \alpha_{\rm eff}^2 (k_B T)^2} {3  n (\hbar v_F)^2}
	\int_0^2dx \, \biggl\{
	e^{-2 d x \sqrt{\pi n \eps_0^\bot / \eps_0^\Vert }}
\nonumber
\\[0.5ex]
	&\frac {x^3 (4 - x^2) } 
			{\bigl[
			(x +  4\alpha_{\rm eff}  )^2 - 16 \alpha_{\rm eff}^2  
			\operatorname{exp} \bigl( -2 d x \sqrt{\pi n \eps_0^\bot / \eps_0^\Vert }  \bigr)
			\bigr]^2}
	\biggr\}
\,
\label{eqn:low_T}
\end{align}
where
$
\alpha_{\rm eff}
	\equiv
	\alpha_g / \sqrt{\eps_0^\bot \eps_0^\Vert}
$
(see Ref.~\onlinecite{Amorim_2012} for details).
%
%
%

\begin{thebibliography}{54}
\expandafter\ifx\csname natexlab\endcsname\relax\def\natexlab#1{#1}\fi
\expandafter\ifx\csname bibnamefont\endcsname\relax
  \def\bibnamefont#1{#1}\fi
\expandafter\ifx\csname bibfnamefont\endcsname\relax
  \def\bibfnamefont#1{#1}\fi
\expandafter\ifx\csname citenamefont\endcsname\relax
  \def\citenamefont#1{#1}\fi
\expandafter\ifx\csname url\endcsname\relax
  \def\url#1{\texttt{#1}}\fi
\expandafter\ifx\csname urlprefix\endcsname\relax\def\urlprefix{URL }\fi
\providecommand{\bibinfo}[2]{#2}
\providecommand{\eprint}[2][]{\url{#2}}

\bibitem[{\citenamefont{Gramila et~al.}(1991)\citenamefont{Gramila, Eisenstein,
  MacDonald, Pfeiffer, and West}}]{Gramila_1991}
\bibinfo{author}{\bibfnamefont{T.~J.} \bibnamefont{Gramila}},
  \bibinfo{author}{\bibfnamefont{J.~P.} \bibnamefont{Eisenstein}},
  \bibinfo{author}{\bibfnamefont{A.~H.} \bibnamefont{MacDonald}},
  \bibinfo{author}{\bibfnamefont{L.~N.} \bibnamefont{Pfeiffer}},
  \bibnamefont{and} \bibinfo{author}{\bibfnamefont{K.~W.} \bibnamefont{West}},
  \bibinfo{journal}{Phys. Rev. Lett.} \textbf{\bibinfo{volume}{66}},
  \bibinfo{pages}{1216} (\bibinfo{year}{1991}).

\bibitem[{\citenamefont{Sivan et~al.}(1992)\citenamefont{Sivan, Solomon, and
  Shtrikman}}]{Sivan_1992}
\bibinfo{author}{\bibfnamefont{U.}~\bibnamefont{Sivan}},
  \bibinfo{author}{\bibfnamefont{P.~M.} \bibnamefont{Solomon}},
  \bibnamefont{and}
  \bibinfo{author}{\bibfnamefont{H.}~\bibnamefont{Shtrikman}},
  \bibinfo{journal}{Phys. Rev. Lett.} \textbf{\bibinfo{volume}{68}},
  \bibinfo{pages}{1196} (\bibinfo{year}{1992}).

\bibitem[{\citenamefont{Zheng and MacDonald}(1993)}]{Zheng_1993}
\bibinfo{author}{\bibfnamefont{L.}~\bibnamefont{Zheng}} \bibnamefont{and}
  \bibinfo{author}{\bibfnamefont{A.~H.} \bibnamefont{MacDonald}},
  \bibinfo{journal}{Phys. Rev. B} \textbf{\bibinfo{volume}{48}},
  \bibinfo{pages}{8203} (\bibinfo{year}{1993}).

\bibitem[{\citenamefont{Flensberg et~al.}(1995)\citenamefont{Flensberg, Hu,
  Jauho, and Kinaret}}]{Flensberg_1995}
\bibinfo{author}{\bibfnamefont{K.}~\bibnamefont{Flensberg}},
  \bibinfo{author}{\bibfnamefont{B.~Y.-K.} \bibnamefont{Hu}},
  \bibinfo{author}{\bibfnamefont{A.-P.} \bibnamefont{Jauho}}, \bibnamefont{and}
  \bibinfo{author}{\bibfnamefont{J.~M.} \bibnamefont{Kinaret}},
  \bibinfo{journal}{Phys. Rev. B} \textbf{\bibinfo{volume}{52}},
  \bibinfo{pages}{14761} (\bibinfo{year}{1995}).

\bibitem[{\citenamefont{Kamenev and Oreg}(1995)}]{Kamenev_1995}
\bibinfo{author}{\bibfnamefont{A.}~\bibnamefont{Kamenev}} \bibnamefont{and}
  \bibinfo{author}{\bibfnamefont{Y.}~\bibnamefont{Oreg}},
  \bibinfo{journal}{Phys. Rev. B} \textbf{\bibinfo{volume}{52}},
  \bibinfo{pages}{7516} (\bibinfo{year}{1995}).

\bibitem[{\citenamefont{Amorim and Peres}(2012)}]{Amorim_2012}
\bibinfo{author}{\bibfnamefont{B.}~\bibnamefont{Amorim}} \bibnamefont{and}
  \bibinfo{author}{\bibfnamefont{N.~M.~R.} \bibnamefont{Peres}},
  \bibinfo{journal}{Journal of Physics: Condensed Matter}
  \textbf{\bibinfo{volume}{24}}, \bibinfo{pages}{335602}
  (\bibinfo{year}{2012}).

\bibitem[{\citenamefont{Carrega et~al.}(2012)\citenamefont{Carrega,
  Tudorovskiy, Principi, Katsnelson, and Polini}}]{Carrega_2012}
\bibinfo{author}{\bibfnamefont{M.}~\bibnamefont{Carrega}},
  \bibinfo{author}{\bibfnamefont{T.}~\bibnamefont{Tudorovskiy}},
  \bibinfo{author}{\bibfnamefont{A.}~\bibnamefont{Principi}},
  \bibinfo{author}{\bibfnamefont{M.~I.} \bibnamefont{Katsnelson}},
  \bibnamefont{and} \bibinfo{author}{\bibfnamefont{M.}~\bibnamefont{Polini}},
  \bibinfo{journal}{New Journal of Physics} \textbf{\bibinfo{volume}{14}},
  \bibinfo{pages}{063033} (\bibinfo{year}{2012}).

\bibitem[{\citenamefont{Katsnelson}(2011)}]{Katsnelson_2011}
\bibinfo{author}{\bibfnamefont{M.~I.} \bibnamefont{Katsnelson}},
  \bibinfo{journal}{Phys. Rev. B} \textbf{\bibinfo{volume}{84}},
  \bibinfo{pages}{041407} (\bibinfo{year}{2011}).

\bibitem[{\citenamefont{Narozhny et~al.}(2012)\citenamefont{Narozhny, Titov,
  Gornyi, and Ostrovsky}}]{Narozhny_2011}
\bibinfo{author}{\bibfnamefont{B.~N.} \bibnamefont{Narozhny}},
  \bibinfo{author}{\bibfnamefont{M.}~\bibnamefont{Titov}},
  \bibinfo{author}{\bibfnamefont{I.~V.} \bibnamefont{Gornyi}},
  \bibnamefont{and} \bibinfo{author}{\bibfnamefont{P.~M.}
  \bibnamefont{Ostrovsky}}, \bibinfo{journal}{Phys. Rev. B}
  \textbf{\bibinfo{volume}{85}}, \bibinfo{pages}{195421}
  (\bibinfo{year}{2012}).

\bibitem[{\citenamefont{Peres et~al.}(2011)\citenamefont{Peres, dos Santos, and
  Neto}}]{Peres_2011b}
\bibinfo{author}{\bibfnamefont{N.~M.~R.} \bibnamefont{Peres}},
  \bibinfo{author}{\bibfnamefont{J.~M. B.~L.} \bibnamefont{dos Santos}},
  \bibnamefont{and} \bibinfo{author}{\bibfnamefont{A.~H.~C.}
  \bibnamefont{Neto}}, \bibinfo{journal}{EPL (Europhysics Letters)}
  \textbf{\bibinfo{volume}{95}}, \bibinfo{pages}{18001} (\bibinfo{year}{2011}).

\bibitem[{\citenamefont{Hwang et~al.}(2011)\citenamefont{Hwang, Sensarma, and
  Das~Sarma}}]{Hwang_2011}
\bibinfo{author}{\bibfnamefont{E.~H.} \bibnamefont{Hwang}},
  \bibinfo{author}{\bibfnamefont{R.}~\bibnamefont{Sensarma}}, \bibnamefont{and}
  \bibinfo{author}{\bibfnamefont{S.}~\bibnamefont{Das~Sarma}},
  \bibinfo{journal}{Phys. Rev. B} \textbf{\bibinfo{volume}{84}},
  \bibinfo{pages}{245441} (\bibinfo{year}{2011}).

\bibitem[{\citenamefont{Tse et~al.}(2007)\citenamefont{Tse, Hu, and
  Das~Sarma}}]{Wang_2007b}
\bibinfo{author}{\bibfnamefont{W.-K.} \bibnamefont{Tse}},
  \bibinfo{author}{\bibfnamefont{B.~Y.-K.} \bibnamefont{Hu}}, \bibnamefont{and}
  \bibinfo{author}{\bibfnamefont{S.}~\bibnamefont{Das~Sarma}},
  \bibinfo{journal}{Phys. Rev. B} \textbf{\bibinfo{volume}{76}},
  \bibinfo{pages}{081401} (\bibinfo{year}{2007}).

\bibitem[{\citenamefont{{Badalyan} and {Peeters}}(2012)}]{Badalyan2012}
\bibinfo{author}{\bibfnamefont{S.~M.} \bibnamefont{{Badalyan}}}
  \bibnamefont{and} \bibinfo{author}{\bibfnamefont{F.~M.}
  \bibnamefont{{Peeters}}}, \bibinfo{journal}{ArXiv e-prints}
  (\bibinfo{year}{2012}), \eprint{1204.4598}.

\bibitem[{\citenamefont{Scharf and Matos-Abiague}(2012)}]{Benedikt2012}
\bibinfo{author}{\bibfnamefont{B.}~\bibnamefont{Scharf}} \bibnamefont{and}
  \bibinfo{author}{\bibfnamefont{A.}~\bibnamefont{Matos-Abiague}},
  \bibinfo{journal}{Phys. Rev. B} \textbf{\bibinfo{volume}{86}},
  \bibinfo{pages}{115425} (\bibinfo{year}{2012}).

\bibitem[{\citenamefont{{Sch{\"u}tt} et~al.}(2012)\citenamefont{{Sch{\"u}tt},
  {Ostrovsky}, {Titov}, {Gornyi}, {Narozhny}, and {Mirlin}}}]{Ostrovski2012b}
\bibinfo{author}{\bibfnamefont{M.}~\bibnamefont{{Sch{\"u}tt}}},
  \bibinfo{author}{\bibfnamefont{P.~M.} \bibnamefont{{Ostrovsky}}},
  \bibinfo{author}{\bibfnamefont{M.}~\bibnamefont{{Titov}}},
  \bibinfo{author}{\bibfnamefont{I.~V.} \bibnamefont{{Gornyi}}},
  \bibinfo{author}{\bibfnamefont{B.~N.} \bibnamefont{{Narozhny}}},
  \bibnamefont{and} \bibinfo{author}{\bibfnamefont{A.~D.}
  \bibnamefont{{Mirlin}}}, \bibinfo{journal}{ArXiv e-prints}
  (\bibinfo{year}{2012}), \eprint{1205.5018}.

\bibitem[{\citenamefont{{Song} and {Levitov}}(2012)}]{Levitov2012}
\bibinfo{author}{\bibfnamefont{J.~C.~W.} \bibnamefont{{Song}}}
  \bibnamefont{and} \bibinfo{author}{\bibfnamefont{L.~S.}
  \bibnamefont{{Levitov}}}, \bibinfo{journal}{ArXiv e-prints}
  (\bibinfo{year}{2012}), \eprint{1205.5257}.

\bibitem[{\citenamefont{Castro~Neto et~al.}(2009)\citenamefont{Castro~Neto,
  Guinea, Peres, Novoselov, and Geim}}]{Guinea_2009}
\bibinfo{author}{\bibfnamefont{A.~H.} \bibnamefont{Castro~Neto}},
  \bibinfo{author}{\bibfnamefont{F.}~\bibnamefont{Guinea}},
  \bibinfo{author}{\bibfnamefont{N.~M.~R.} \bibnamefont{Peres}},
  \bibinfo{author}{\bibfnamefont{K.~S.} \bibnamefont{Novoselov}},
  \bibnamefont{and} \bibinfo{author}{\bibfnamefont{A.~K.} \bibnamefont{Geim}},
  \bibinfo{journal}{Rev. Mod. Phys.} \textbf{\bibinfo{volume}{81}},
  \bibinfo{pages}{109} (\bibinfo{year}{2009}).

\bibitem[{\citenamefont{Kim et~al.}(2011)\citenamefont{Kim, Jo, Nah, Yao,
  Banerjee, and Tutuc}}]{Kim_2011}
\bibinfo{author}{\bibfnamefont{S.}~\bibnamefont{Kim}},
  \bibinfo{author}{\bibfnamefont{I.}~\bibnamefont{Jo}},
  \bibinfo{author}{\bibfnamefont{J.}~\bibnamefont{Nah}},
  \bibinfo{author}{\bibfnamefont{Z.}~\bibnamefont{Yao}},
  \bibinfo{author}{\bibfnamefont{S.~K.} \bibnamefont{Banerjee}},
  \bibnamefont{and} \bibinfo{author}{\bibfnamefont{E.}~\bibnamefont{Tutuc}},
  \bibinfo{journal}{Phys. Rev. B} \textbf{\bibinfo{volume}{83}},
  \bibinfo{pages}{161401} (\bibinfo{year}{2011}).

\bibitem[{\citenamefont{Kim and Tutuc}(2012)}]{Kim_2012}
\bibinfo{author}{\bibfnamefont{S.}~\bibnamefont{Kim}} \bibnamefont{and}
  \bibinfo{author}{\bibfnamefont{E.}~\bibnamefont{Tutuc}},
  \bibinfo{journal}{Solid State Communications} \textbf{\bibinfo{volume}{152}},
  \bibinfo{pages}{1283 } (\bibinfo{year}{2012}).

\bibitem[{\citenamefont{{Gorbachev} et~al.}(2012)\citenamefont{{Gorbachev},
  {Geim}, {Katsnelson}, {Novoselov}, {Tudorovskiy}, {Grigorieva}, {MacDonald},
  {Watanabe}, {Taniguchi}, and {Ponomarenko}}}]{Gorbachev_2012}
\bibinfo{author}{\bibfnamefont{R.~V.} \bibnamefont{{Gorbachev}}},
  \bibinfo{author}{\bibfnamefont{A.~K.} \bibnamefont{{Geim}}},
  \bibinfo{author}{\bibfnamefont{M.~I.} \bibnamefont{{Katsnelson}}},
  \bibinfo{author}{\bibfnamefont{K.~S.} \bibnamefont{{Novoselov}}},
  \bibinfo{author}{\bibfnamefont{T.}~\bibnamefont{{Tudorovskiy}}},
  \bibinfo{author}{\bibfnamefont{I.~V.} \bibnamefont{{Grigorieva}}},
  \bibinfo{author}{\bibfnamefont{A.~H.} \bibnamefont{{MacDonald}}},
  \bibinfo{author}{\bibfnamefont{K.}~\bibnamefont{{Watanabe}}},
  \bibinfo{author}{\bibfnamefont{T.}~\bibnamefont{{Taniguchi}}},
  \bibnamefont{and} \bibinfo{author}{\bibfnamefont{L.~A.}
  \bibnamefont{{Ponomarenko}}}, \bibinfo{journal}{ArXiv e-prints}
  (\bibinfo{year}{2012}), \eprint{1206.6626}.

\bibitem[{\citenamefont{Jauho and Smith}(1993)}]{Jauho_1993}
\bibinfo{author}{\bibfnamefont{A.-P.} \bibnamefont{Jauho}} \bibnamefont{and}
  \bibinfo{author}{\bibfnamefont{H.}~\bibnamefont{Smith}},
  \bibinfo{journal}{Phys. Rev. B} \textbf{\bibinfo{volume}{47}},
  \bibinfo{pages}{4420} (\bibinfo{year}{1993}).

\bibitem[{\citenamefont{Dean et~al.}(2010)\citenamefont{Dean, Young, Meric,
  Lee, Wang, Sorgenfrei, Watanabe, Taniguchi, Kim, Shepard et~al.}}]{Dean_2010}
\bibinfo{author}{\bibfnamefont{R.~C.} \bibnamefont{Dean}},
  \bibinfo{author}{\bibfnamefont{A.~F.} \bibnamefont{Young}},
  \bibinfo{author}{\bibfnamefont{I.}~\bibnamefont{Meric}},
  \bibinfo{author}{\bibfnamefont{C.}~\bibnamefont{Lee}},
  \bibinfo{author}{\bibfnamefont{L.}~\bibnamefont{Wang}},
  \bibinfo{author}{\bibfnamefont{S.}~\bibnamefont{Sorgenfrei}},
  \bibinfo{author}{\bibfnamefont{K.}~\bibnamefont{Watanabe}},
  \bibinfo{author}{\bibfnamefont{T.}~\bibnamefont{Taniguchi}},
  \bibinfo{author}{\bibfnamefont{P.}~\bibnamefont{Kim}},
  \bibinfo{author}{\bibfnamefont{K.~L.} \bibnamefont{Shepard}},
  \bibnamefont{et~al.}, \bibinfo{journal}{Nat Nano}
  \textbf{\bibinfo{volume}{5}}, \bibinfo{pages}{722} (\bibinfo{year}{2010}).

\bibitem[{\citenamefont{Mayorov et~al.}(2011)\citenamefont{Mayorov, Gorbachev,
  Morozov, Britnell, Jalil, Ponomarenko, Blake, Novoselov, Watanabe, Taniguchi
  et~al.}}]{Mayorov_2011}
\bibinfo{author}{\bibfnamefont{A.~S.} \bibnamefont{Mayorov}},
  \bibinfo{author}{\bibfnamefont{R.~V.} \bibnamefont{Gorbachev}},
  \bibinfo{author}{\bibfnamefont{S.~V.} \bibnamefont{Morozov}},
  \bibinfo{author}{\bibfnamefont{L.}~\bibnamefont{Britnell}},
  \bibinfo{author}{\bibfnamefont{R.}~\bibnamefont{Jalil}},
  \bibinfo{author}{\bibfnamefont{L.~A.} \bibnamefont{Ponomarenko}},
  \bibinfo{author}{\bibfnamefont{P.}~\bibnamefont{Blake}},
  \bibinfo{author}{\bibfnamefont{K.~S.} \bibnamefont{Novoselov}},
  \bibinfo{author}{\bibfnamefont{K.}~\bibnamefont{Watanabe}},
  \bibinfo{author}{\bibfnamefont{T.}~\bibnamefont{Taniguchi}},
  \bibnamefont{et~al.}, \bibinfo{journal}{Nano Letters}
  \textbf{\bibinfo{volume}{11}}, \bibinfo{pages}{2396} (\bibinfo{year}{2011}).

\bibitem[{\citenamefont{Schiefele et~al.}(2012)\citenamefont{Schiefele, Sols,
  and Guinea}}]{Schiefele_2012}
\bibinfo{author}{\bibfnamefont{J.}~\bibnamefont{Schiefele}},
  \bibinfo{author}{\bibfnamefont{F.}~\bibnamefont{Sols}}, \bibnamefont{and}
  \bibinfo{author}{\bibfnamefont{F.}~\bibnamefont{Guinea}},
  \bibinfo{journal}{Phys. Rev. B} \textbf{\bibinfo{volume}{85}},
  \bibinfo{pages}{195420} (\bibinfo{year}{2012}).

\bibitem[{\citenamefont{Garcia et~al.}(2012)\citenamefont{Garcia, Wurstbauer,
  Levy, Pfeiffer, Pinczuk, Plaut, Wang, Dean, Buizza, Zande
  et~al.}}]{Garcia_2012}
\bibinfo{author}{\bibfnamefont{J.~M.} \bibnamefont{Garcia}},
  \bibinfo{author}{\bibfnamefont{U.}~\bibnamefont{Wurstbauer}},
  \bibinfo{author}{\bibfnamefont{A.}~\bibnamefont{Levy}},
  \bibinfo{author}{\bibfnamefont{L.~N.} \bibnamefont{Pfeiffer}},
  \bibinfo{author}{\bibfnamefont{A.}~\bibnamefont{Pinczuk}},
  \bibinfo{author}{\bibfnamefont{A.~S.} \bibnamefont{Plaut}},
  \bibinfo{author}{\bibfnamefont{L.}~\bibnamefont{Wang}},
  \bibinfo{author}{\bibfnamefont{C.~R.} \bibnamefont{Dean}},
  \bibinfo{author}{\bibfnamefont{R.}~\bibnamefont{Buizza}},
  \bibinfo{author}{\bibfnamefont{A.~V.~D.} \bibnamefont{Zande}},
  \bibnamefont{et~al.}, \bibinfo{journal}{Solid State Communications}
  \textbf{\bibinfo{volume}{152}}, \bibinfo{pages}{975} (\bibinfo{year}{2012}).

\bibitem[{\citenamefont{Ponomarenko et~al.}(2011)\citenamefont{Ponomarenko,
  Geim, Zhukov, Jalil, Morozov, Novoselov, Grigorieva, Hill, Cheianov, Fal/'ko
  et~al.}}]{Ponomarenko_2011}
\bibinfo{author}{\bibfnamefont{L.~A.} \bibnamefont{Ponomarenko}},
  \bibinfo{author}{\bibfnamefont{A.~K.} \bibnamefont{Geim}},
  \bibinfo{author}{\bibfnamefont{A.~A.} \bibnamefont{Zhukov}},
  \bibinfo{author}{\bibfnamefont{R.}~\bibnamefont{Jalil}},
  \bibinfo{author}{\bibfnamefont{S.~V.} \bibnamefont{Morozov}},
  \bibinfo{author}{\bibfnamefont{K.~S.} \bibnamefont{Novoselov}},
  \bibinfo{author}{\bibfnamefont{I.~V.} \bibnamefont{Grigorieva}},
  \bibinfo{author}{\bibfnamefont{E.~H.} \bibnamefont{Hill}},
  \bibinfo{author}{\bibfnamefont{V.~V.} \bibnamefont{Cheianov}},
  \bibinfo{author}{\bibfnamefont{V.~I.} \bibnamefont{Fal/'ko}},
  \bibnamefont{et~al.}, \bibinfo{journal}{Nat Phys}
  \textbf{\bibinfo{volume}{7}}, \bibinfo{pages}{958} (\bibinfo{year}{2011}).

\bibitem[{\citenamefont{Britnell
  et~al.}(2012{\natexlab{a}})\citenamefont{Britnell, Gorbachev, Jalil, Belle,
  Schedin, Mishchenko, Georgiou, Katsnelson, Eaves, Morozov
  et~al.}}]{Britnell_2012}
\bibinfo{author}{\bibfnamefont{L.}~\bibnamefont{Britnell}},
  \bibinfo{author}{\bibfnamefont{R.~V.} \bibnamefont{Gorbachev}},
  \bibinfo{author}{\bibfnamefont{R.}~\bibnamefont{Jalil}},
  \bibinfo{author}{\bibfnamefont{B.~D.} \bibnamefont{Belle}},
  \bibinfo{author}{\bibfnamefont{F.}~\bibnamefont{Schedin}},
  \bibinfo{author}{\bibfnamefont{A.}~\bibnamefont{Mishchenko}},
  \bibinfo{author}{\bibfnamefont{T.}~\bibnamefont{Georgiou}},
  \bibinfo{author}{\bibfnamefont{M.~I.} \bibnamefont{Katsnelson}},
  \bibinfo{author}{\bibfnamefont{L.}~\bibnamefont{Eaves}},
  \bibinfo{author}{\bibfnamefont{S.~V.} \bibnamefont{Morozov}},
  \bibnamefont{et~al.}, \bibinfo{journal}{Science}
  \textbf{\bibinfo{volume}{335}}, \bibinfo{pages}{947}
  (\bibinfo{year}{2012}{\natexlab{a}}).

\bibitem[{\citenamefont{Britnell
  et~al.}(2012{\natexlab{b}})\citenamefont{Britnell, Gorbachev, Jalil, Belle,
  Schedin, Katsnelson, Eaves, Morozov, Mayorov, Peres et~al.}}]{Britnell_2012b}
\bibinfo{author}{\bibfnamefont{L.}~\bibnamefont{Britnell}},
  \bibinfo{author}{\bibfnamefont{R.~V.} \bibnamefont{Gorbachev}},
  \bibinfo{author}{\bibfnamefont{R.}~\bibnamefont{Jalil}},
  \bibinfo{author}{\bibfnamefont{B.~D.} \bibnamefont{Belle}},
  \bibinfo{author}{\bibfnamefont{F.}~\bibnamefont{Schedin}},
  \bibinfo{author}{\bibfnamefont{M.~I.} \bibnamefont{Katsnelson}},
  \bibinfo{author}{\bibfnamefont{L.}~\bibnamefont{Eaves}},
  \bibinfo{author}{\bibfnamefont{S.~V.} \bibnamefont{Morozov}},
  \bibinfo{author}{\bibfnamefont{A.~S.} \bibnamefont{Mayorov}},
  \bibinfo{author}{\bibfnamefont{N.~M.~R.} \bibnamefont{Peres}},
  \bibnamefont{et~al.}, \bibinfo{journal}{Nano Letters}
  \textbf{\bibinfo{volume}{12}}, \bibinfo{pages}{1707}
  (\bibinfo{year}{2012}{\natexlab{b}}).

\bibitem[{\citenamefont{Jalabert and Das~Sarma}(1989)}]{Jalabert_1989}
\bibinfo{author}{\bibfnamefont{R.}~\bibnamefont{Jalabert}} \bibnamefont{and}
  \bibinfo{author}{\bibfnamefont{S.}~\bibnamefont{Das~Sarma}},
  \bibinfo{journal}{Phys. Rev. B} \textbf{\bibinfo{volume}{40}},
  \bibinfo{pages}{9723} (\bibinfo{year}{1989}).

\bibitem[{\citenamefont{Tso et~al.}(1992)\citenamefont{Tso, Vasilopoulos, and
  Peeters}}]{Tso_1992}
\bibinfo{author}{\bibfnamefont{H.~C.} \bibnamefont{Tso}},
  \bibinfo{author}{\bibfnamefont{P.}~\bibnamefont{Vasilopoulos}},
  \bibnamefont{and} \bibinfo{author}{\bibfnamefont{F.~M.}
  \bibnamefont{Peeters}}, \bibinfo{journal}{Phys. Rev. Lett.}
  \textbf{\bibinfo{volume}{68}}, \bibinfo{pages}{2516} (\bibinfo{year}{1992}).

\bibitem[{\citenamefont{Zhang and Takahashi}(1993)}]{Zhang_1993c}
\bibinfo{author}{\bibfnamefont{C.}~\bibnamefont{Zhang}} \bibnamefont{and}
  \bibinfo{author}{\bibfnamefont{Y.}~\bibnamefont{Takahashi}},
  \bibinfo{journal}{Journal of Physics: Condensed Matter}
  \textbf{\bibinfo{volume}{5}}, \bibinfo{pages}{5009} (\bibinfo{year}{1993}).

\bibitem[{\citenamefont{Gramila et~al.}(1993)\citenamefont{Gramila, Eisenstein,
  MacDonald, Pfeiffer, and West}}]{Gramila_1993}
\bibinfo{author}{\bibfnamefont{T.~J.} \bibnamefont{Gramila}},
  \bibinfo{author}{\bibfnamefont{J.~P.} \bibnamefont{Eisenstein}},
  \bibinfo{author}{\bibfnamefont{A.~H.} \bibnamefont{MacDonald}},
  \bibinfo{author}{\bibfnamefont{L.~N.} \bibnamefont{Pfeiffer}},
  \bibnamefont{and} \bibinfo{author}{\bibfnamefont{K.~W.} \bibnamefont{West}},
  \bibinfo{journal}{Phys. Rev. B} \textbf{\bibinfo{volume}{47}},
  \bibinfo{pages}{12957} (\bibinfo{year}{1993}).

\bibitem[{\citenamefont{G\"uven and Tanatar}(1997)}]{Guven_1997}
\bibinfo{author}{\bibfnamefont{K.}~\bibnamefont{G\"uven}} \bibnamefont{and}
  \bibinfo{author}{\bibfnamefont{B.}~\bibnamefont{Tanatar}},
  \bibinfo{journal}{Phys. Rev. B} \textbf{\bibinfo{volume}{56}},
  \bibinfo{pages}{7535} (\bibinfo{year}{1997}).

\bibitem[{\citenamefont{B\o{}nsager et~al.}(1998)\citenamefont{B\o{}nsager,
  Flensberg, Yu-Kuang~Hu, and MacDonald}}]{Bonsager_1998}
\bibinfo{author}{\bibfnamefont{M.~C.} \bibnamefont{B\o{}nsager}},
  \bibinfo{author}{\bibfnamefont{K.}~\bibnamefont{Flensberg}},
  \bibinfo{author}{\bibfnamefont{B.}~\bibnamefont{Yu-Kuang~Hu}},
  \bibnamefont{and} \bibinfo{author}{\bibfnamefont{A.~H.}
  \bibnamefont{MacDonald}}, \bibinfo{journal}{Phys. Rev. B}
  \textbf{\bibinfo{volume}{57}}, \bibinfo{pages}{7085} (\bibinfo{year}{1998}).

\bibitem[{\citenamefont{Fr{\"o}hlich}(1954)}]{Frohlich_1954}
\bibinfo{author}{\bibfnamefont{H.}~\bibnamefont{Fr{\"o}hlich}},
  \bibinfo{journal}{Advances in Physics} \textbf{\bibinfo{volume}{3}},
  \bibinfo{pages}{325} (\bibinfo{year}{1954}).

\bibitem[{\citenamefont{Mahan}(1981)}]{Mahan_old}
\bibinfo{author}{\bibfnamefont{G.~D.} \bibnamefont{Mahan}},
  \emph{\bibinfo{title}{Many-particle physics}} (\bibinfo{publisher}{Plenum
  Press}, \bibinfo{address}{New York}, \bibinfo{year}{1981}).

\bibitem[{\citenamefont{Marder}(2010)}]{Marder}
\bibinfo{author}{\bibfnamefont{M.~P.} \bibnamefont{Marder}},
  \emph{\bibinfo{title}{Condensed Matter Physics}} (\bibinfo{publisher}{John
  Wiley \& Sons, Inc.}, \bibinfo{address}{New Jersey}, \bibinfo{year}{2010}),
  \bibinfo{edition}{2nd} ed.

\bibitem[{\citenamefont{Fratini and Guinea}(2008)}]{Fratini_2008}
\bibinfo{author}{\bibfnamefont{S.}~\bibnamefont{Fratini}} \bibnamefont{and}
  \bibinfo{author}{\bibfnamefont{F.}~\bibnamefont{Guinea}},
  \bibinfo{journal}{Phys. Rev. B} \textbf{\bibinfo{volume}{77}},
  \bibinfo{pages}{195415} (\bibinfo{year}{2008}).

\bibitem[{\citenamefont{Chen et~al.}(2008)\citenamefont{Chen, Jang, Xiao,
  Ishigami, and Fuhrer}}]{Chen_2008}
\bibinfo{author}{\bibfnamefont{J.-H.} \bibnamefont{Chen}},
  \bibinfo{author}{\bibfnamefont{C.}~\bibnamefont{Jang}},
  \bibinfo{author}{\bibfnamefont{S.}~\bibnamefont{Xiao}},
  \bibinfo{author}{\bibfnamefont{M.}~\bibnamefont{Ishigami}}, \bibnamefont{and}
  \bibinfo{author}{\bibfnamefont{M.~S.} \bibnamefont{Fuhrer}},
  \bibinfo{journal}{Nature Nanotechnology} \textbf{\bibinfo{volume}{3}},
  \bibinfo{pages}{206} (\bibinfo{year}{2008}).

\bibitem[{Note1()}]{Note1}
\bibinfo{note}{See Refs.~\protect \onlinecite {Geick_1966, Michel_2011a,
  Serrano_2007} for details on the phonon dispersions of hBN, and the
  classification of the vibrational modes into Raman active, infrared active
  and optically silent. Figure~3 and eqns.~(21) and (24) of Ref.~\protect
  \onlinecite {Michel_2011a} show how the long range Coulomb potential
  associated with the infrared active modes leads to the splitting of
  transverse and longitudinal optical frequencies at the $\Gamma $ point.}

\bibitem[{\citenamefont{Loudon}(1964)}]{Loudon}
\bibinfo{author}{\bibfnamefont{R.}~\bibnamefont{Loudon}},
  \bibinfo{journal}{Advances in Physics} \textbf{\bibinfo{volume}{13}},
  \bibinfo{pages}{423} (\bibinfo{year}{1964}).

\bibitem[{Note2()}]{Note2}
\bibinfo{note}{We are here using the retarded expression (defined as
  being analytic in the upper half of the complex $\omega $ plane) in order to
  be consistent with the likewise retarded polarizability of graphene taken
  from Ref.~\protect \onlinecite {Wunsch_2006}. Not keeping this consistency
  yields significantly different results.}

\bibitem[{\citenamefont{Sarma and Mason}(1985)}]{Das_Sarma_1985}
\bibinfo{author}{\bibfnamefont{S.}~\bibnamefont{Sarma}} \bibnamefont{and}
  \bibinfo{author}{\bibfnamefont{B.}~\bibnamefont{Mason}},
  \bibinfo{journal}{Annals of Physics} \textbf{\bibinfo{volume}{163}},
  \bibinfo{pages}{78 } (\bibinfo{year}{1985}).

\bibitem[{\citenamefont{Geick et~al.}(1966)\citenamefont{Geick, Perry, and
  Rupprecht}}]{Geick_1966}
\bibinfo{author}{\bibfnamefont{R.}~\bibnamefont{Geick}},
  \bibinfo{author}{\bibfnamefont{C.~H.} \bibnamefont{Perry}}, \bibnamefont{and}
  \bibinfo{author}{\bibfnamefont{G.}~\bibnamefont{Rupprecht}},
  \bibinfo{journal}{Phys. Rev.} \textbf{\bibinfo{volume}{146}},
  \bibinfo{pages}{543} (\bibinfo{year}{1966}).

\bibitem[{Note3()}]{Note3}
\bibinfo{note}{In the numerical calculations, we use for simplicity the
  zero temperature expression for $\chi $ as calculated in Refs.~\protect
  \onlinecite {Wunsch_2006,Hwang_2007}, which is a good approximation for $T
  \ll T_F$, with $T_F$ the Fermi temperature.}

\bibitem[{\citenamefont{Lyddane et~al.}(1941)\citenamefont{Lyddane, Sachs, and
  Teller}}]{Lyddane_1941}
\bibinfo{author}{\bibfnamefont{R.~H.} \bibnamefont{Lyddane}},
  \bibinfo{author}{\bibfnamefont{R.~G.} \bibnamefont{Sachs}}, \bibnamefont{and}
  \bibinfo{author}{\bibfnamefont{E.}~\bibnamefont{Teller}},
  \bibinfo{journal}{Phys. Rev.} \textbf{\bibinfo{volume}{59}},
  \bibinfo{pages}{673} (\bibinfo{year}{1941}).

\bibitem[{Note4()}]{Note4}
\bibinfo{note}{We here use a simplified form of the nonlinear
  susceptibility of graphene, which is valid for electron doping high enough
  such that the existence of the valence band can be ignored. The condition
  $T\ll T_F$ is important as we use the zero temperature expressions for the
  polarizability of graphene. See Refs.~\protect \onlinecite
  {Peres_2011b,Amorim_2012} for a discussion of both approximations.}

\bibitem[{Note5()}]{Note5}
\bibinfo{note}{See Fig.~2a in Ref.~\protect \onlinecite
  {Gorbachev_2012}.}

\bibitem[{\citenamefont{Stroscio and Dutta}(2003)}]{Stroscio}
\bibinfo{author}{\bibfnamefont{M.~A.} \bibnamefont{Stroscio}} \bibnamefont{and}
  \bibinfo{author}{\bibfnamefont{M.}~\bibnamefont{Dutta}},
  \emph{\bibinfo{title}{Phonons in Nanostructures}}
  (\bibinfo{publisher}{Cambridge University Press},
  \bibinfo{address}{Cambridge}, \bibinfo{year}{2003}).

\bibitem[{\citenamefont{Mori and Ando}(1989)}]{Mori}
\bibinfo{author}{\bibfnamefont{N.}~\bibnamefont{Mori}} \bibnamefont{and}
  \bibinfo{author}{\bibfnamefont{T.}~\bibnamefont{Ando}},
  \bibinfo{journal}{Phys. Rev. B} \textbf{\bibinfo{volume}{40}},
  \bibinfo{pages}{6175} (\bibinfo{year}{1989}).

\bibitem[{\citenamefont{Michel and Verberck}(2011)}]{Michel_2011a}
\bibinfo{author}{\bibfnamefont{K.~H.} \bibnamefont{Michel}} \bibnamefont{and}
  \bibinfo{author}{\bibfnamefont{B.}~\bibnamefont{Verberck}},
  \bibinfo{journal}{Phys. Rev. B} \textbf{\bibinfo{volume}{83}},
  \bibinfo{pages}{115328} (\bibinfo{year}{2011}).

\bibitem[{\citenamefont{Serrano et~al.}(2007)\citenamefont{Serrano, Bosak,
  Arenal, Krisch, Watanabe, Taniguchi, Kanda, Rubio, and Wirtz}}]{Serrano_2007}
\bibinfo{author}{\bibfnamefont{J.}~\bibnamefont{Serrano}},
  \bibinfo{author}{\bibfnamefont{A.}~\bibnamefont{Bosak}},
  \bibinfo{author}{\bibfnamefont{R.}~\bibnamefont{Arenal}},
  \bibinfo{author}{\bibfnamefont{M.}~\bibnamefont{Krisch}},
  \bibinfo{author}{\bibfnamefont{K.}~\bibnamefont{Watanabe}},
  \bibinfo{author}{\bibfnamefont{T.}~\bibnamefont{Taniguchi}},
  \bibinfo{author}{\bibfnamefont{H.}~\bibnamefont{Kanda}},
  \bibinfo{author}{\bibfnamefont{A.}~\bibnamefont{Rubio}}, \bibnamefont{and}
  \bibinfo{author}{\bibfnamefont{L.}~\bibnamefont{Wirtz}},
  \bibinfo{journal}{Phys. Rev. Lett.} \textbf{\bibinfo{volume}{98}},
  \bibinfo{pages}{095503} (\bibinfo{year}{2007}).

\bibitem[{\citenamefont{Wunsch et~al.}(2006)\citenamefont{Wunsch, Stauber,
  Sols, and Guinea}}]{Wunsch_2006}
\bibinfo{author}{\bibfnamefont{B.}~\bibnamefont{Wunsch}},
  \bibinfo{author}{\bibfnamefont{T.}~\bibnamefont{Stauber}},
  \bibinfo{author}{\bibfnamefont{F.}~\bibnamefont{Sols}}, \bibnamefont{and}
  \bibinfo{author}{\bibfnamefont{F.}~\bibnamefont{Guinea}},
  \bibinfo{journal}{New Journal of Physics} \textbf{\bibinfo{volume}{8}},
  \bibinfo{pages}{318} (\bibinfo{year}{2006}).

\bibitem[{\citenamefont{Hwang and Das~Sarma}(2007)}]{Hwang_2007}
\bibinfo{author}{\bibfnamefont{E.~H.} \bibnamefont{Hwang}} \bibnamefont{and}
  \bibinfo{author}{\bibfnamefont{S.}~\bibnamefont{Das~Sarma}},
  \bibinfo{journal}{Phys. Rev. B} \textbf{\bibinfo{volume}{75}},
  \bibinfo{pages}{205418} (\bibinfo{year}{2007}).

\end{thebibliography}
%
%

%
%
%
%
%
%
\end{document}